\documentclass[preprint,12pt]{elsarticle}
\usepackage{ifpdf}
\usepackage{graphicx}
\usepackage{amssymb}
\usepackage{amsmath}
\biboptions{sort&compress}
\journal{Communications in Nonlinear Science and Numerical Simulation}
\begin{document}

\begin{frontmatter} 

\title{On the symmetries of a nonlinear non-polynomial oscillator}
\author[bdu]{R. Mohanasubha}
\author[bdu]{M. Senthilvelan\corref{cor1}}
\ead{velan@cnld.bdu.ac.in}

\address[bdu]{Centre for Nonlinear Dynamics, School of Physics, Bharathidasan University, Tiruchirappalli - 620 024, India}
\cortext[cor1]{Corresponding author} 





\begin{abstract}
In this paper, we unearth symmetries of different types of a nonlinear non-polynomial oscillator. The symmetries which we report here are adjoint-symmetries, contact symmetries and telescopic vector fields. We also obtain Jacobi last multipliers and Darboux polynomials as a by-product of our procedure. All the aforementioned quantities are derived from a Theorem proved by Muriel and Romero. The procedure which we present here is applicable to a class of nonlinear oscillator equations.
\end{abstract}


\end{frontmatter}


\section{Introduction}
Recently, the complete integrability of the nonlinear oscillator equation
\begin{equation}
\ddot x-\frac{kx}{(1+kx^2)}{\dot x}^2+\frac{\alpha^2x}{(1+kx^2)}=0,
\label{eq1}
\end{equation}
where $k$ and $\alpha$ are arbitrary parameters and overdot denotes the differentiation with respect to $t$, has been demonstrated through $\lambda$-symmetry approach \cite{bhu_ml}. Eq.(\ref{eq1}) admits the general solution of the form $x(t)= A \sin(\Omega t + \delta)$, where $A$ and $\delta$ are two arbitrary constants and the frequency $\Omega$ is related to the amplitude $A$ through the relation $\Omega^2=\frac{k^2}{1+kA^2}$. Eq.(\ref{eq1}) was introduced by Mathews and Lakshmanan in 1974 \cite{ml11,ml22,ml33}. Later Higgs and Leeman studied the classical and quantum dynamics of the oscillator (\ref{eq1}) on the spherical configuration space \cite{higgs,leemon}.

During the past few years a considerable number of studies have been devoted to analyze the physical and mathematical properties of Eq.(\ref{eq1}) \cite{sir1,recent,nuc2,new1,new11}. 
Eq.(\ref{eq1}) is obtainable from the Lagrangian $L=\frac{1}{2}\big(\frac{\dot{x}-\alpha^2x^2}{1+kx^2}\big)$ from which the Hamiltonian can be given in the form $H=\frac{1}{2}p^2(1+k x^2)+\frac{1}{2}\frac{\alpha^2x^2}{1+k x^2}$, where $p=\frac{\dot{x}}{1+kx^2}$. One might note that both the Lagrangian and the Hamiltonian are given in ``non-standard form" \cite{anna1}. In fact the Hamiltonian given above is now being called ``position-dependent mass Hamiltonian''\cite{new2,new3,ano_new2,new4}. The term $(1+kx^2)$ which appear along with momentum in the Hamiltonian can be treated as mass, that is $m(x)=\frac{1}{1+kx^2}$. In the contemporary literature, the quantization of position-dependent mass Hamiltonian is of considerable interest \cite{cari_2,ruby,qunt1,cari_1,lopez,ano_new1}. 
Very recently, the classical and quantum dynamics of the nonlinear non-polynomial oscillator  (\ref{eq1}) and its higher dimensional versions have been studied in various aspects, see for example Refs.\cite{66,1,2,3,4,5}. Eq.(\ref{eq1}) is integrable by quadrature but admits only time-translational symmetry \cite{ml11}.

Since Eq.(\ref{eq1}) lacks sufficient number of Lie point symmetries, the complete integrability of it from symmetries perspective was achieved through $\lambda$-symmetry approach \cite{bhu_ml}. The aim of this paper is to extend the earlier result \cite{bhu_ml} and to demonstrate that the above nonlinear non-polynomial oscillator equation does possess other symmetries, namely adjoint symmetries, contact symmetries, telescopic vector fields and other integrability quantifiers such as Jacobi last multipliers and Darboux polynomials. An important framework which we make here is that one can determine all the aforementioned quantities from $\lambda$-symmetry approach itself \cite{mur}. Through this demonstration we give not only new dimensions to the $\lambda$-symmetry approach but also report additional symmetries exhibited by Eq.(\ref{eq1}). The approach introduced here is applicable to a class of nonlinear oscillators.

We extract all our results from Theorem 5 and the Corollary 6 given by Muriel and Romero \cite{MR09}. To be self-contained, in the following, we recall this corollary: `` If $\lambda$ is such that $v=\frac{\partial} {\partial x}$ is a $\lambda$-symmetry of (\ref{eq1}) the any solution $\mu$ of the first-order linear system $
D[\mu]+(\phi_{\dot{x}}-\lambda)\mu=0,~~\mu_x+(\lambda \mu)_{\dot{x}}=0$
is an integrating factor of (\ref{eq1})''.  Here the function $\mu$ should satisfy both the equations in order to be an integrating factor for the given second-order ordinary differential equation (ODE) $\ddot{x}=\phi(t,x,\dot{x})$. Since Eq.(\ref{eq1}) admits only time-translational symmetry, in order to establish the integrability of Eq. (\ref{eq1}) through symmetry perspective, one should probe other symmetries, namely $\lambda$-symmetries, telescopic vector fields and contact symmetries. Even a symmetry vector field is determined, finding the associated integral/integrating factor is often difficult. Overcoming this obstacle, the Theorem 5 and the Corollary 6 given in Ref. \cite{MR09} describes a procedure to derive integrating factors from $\lambda$-symmetries. In this respect, they can be treated as important. However,  in this work, we utilize the aforementioned Corollary to identify the different types of symmetries of (\ref{eq1}). 

 In our analysis, we consider two cases, namely (i) $\lambda=0$ and (ii) $\lambda \neq 0$. In the first case we have $D[\mu]+\phi_{\dot{x}}\mu=0$. The reduced equation is nothing but the Jacobi last multiplier determining equation for the given dynamical system. Solving this equation we can obtain the Jacobi last multiplier $\mu(=M)$. The obtained $\mu$ need not satisfy the second expression, that is $\mu_x=0$, since $\mu$ is not an integrating factor. Once Jacobi last multiplier is known the inverse of it readily provides Darboux polynomials \cite{suba1}. From Jacobi last multiplier we can derive the Lagrangian for the given equation. In the second case, we consider $\lambda$ is a non-zero function and is determined already for the given equation. Now solving the first equation, that is $D[\mu]+(\phi_{\dot{x}}-\lambda)\mu=0$, we can obtain $\mu$. The obtained $\mu$ may or may not satisfy the second equation, that is $\mu_x+(\lambda \mu)_{\dot{x}}=0$. If it satisfies then it acts as an adjoint-symmetry as well as integrating factor for the given equation. In the other case it just acts as an adjoint-symmetry for the given ODE. In this sense the adjoint-symmetries can be identified from $\lambda$-symmetry approach. To determine contact symmetries we recall the relation \cite{Mur4} $\lambda=\frac{D[Q]}{Q}$, where $Q(=\eta-\dot{x}\xi)$ is the characteristics and $D$ is the total differential operator, that is $D=\frac{\partial}{\partial t}+\dot{x}\frac{\partial}{\partial x}+\phi \frac{\partial}{\partial \dot{x}}$. With known $\lambda$, we solve this equation and obtain $Q$ from which we identify the functions $\xi$ and $\eta$ which in turn gives us either a contact symmetry or a point symmetry for the considered equation. To obtain telescopic vector fields, we recall the relation, \cite{Mur4} which connects the telescopic vector field with $\lambda$-symmetries, that is $\lambda=\frac{\eta^{[\lambda,(1)]}-\phi\xi}{\eta-\dot{x} \xi}$, where $\eta^{[\lambda,(1)]}$ is the first prolongation whose explicit expression is given in (\ref{tel_fir_pro}). This expression contains three unknowns ($\xi,~\eta,~\eta^{[\lambda,(1)]}$). These three functions compose the telescopic vector field for the given $\lambda$. To determine these three unknowns we assume an ansatz for $\xi,~\eta$ and $\eta^{[\lambda,(1)]}$. We choose $\xi=0$  so that the above expression reduces to $\lambda=\frac{\eta^{[\lambda,(1)]}}{\eta}$. We solve this reduced equation and determine the other two components, $\eta$ and $\eta^{[\lambda,(1)]}$. Thus in the second case ($\lambda \neq 0$) we obtain point/contact symmetries and telescopic vector fields for the given ODE from the known $\lambda$-function.


We organize our work as follows. In Sec. II, we briefly recall the method of finding telescopic vector fields for a second-order ODE. In Sec. III, we outline the procedure to obtain contact symmetries and telescopic vector fields from the $\lambda$-symmetry approach. We also obtain Darboux polynomials, Jacobi last multiplier and the Lagrangian associated with the given ODE from the Corollary given by Muriel and Romero \cite{MR09}. In Sec. IV, we illustrate our procedure for the nonlinear non-polynomial oscillator Eq.(\ref{eq1}) and obtain the aforementioned quantities. In Sec. V, we present our conclusions. 

\section{Telescopic vector fields}
 In this section, we briefly recall the method of finding telescopic vector fields for a second-order nonlinear ODE.
 A telescopic vector field can be considered as the $\lambda$-prolongation of contact symmetries since the functions $\xi$ and $\eta$ are allowed to depend on the first derivative of the dependent variable \cite{pucci}. If we put $\lambda=0$ in the telescopic vector field prolongation formula,  we end up at the contact symmetry prolongation formula \cite{Mur4}.  However, unlike Lie point symmetries these telescopic vector fields cannot be determined so easily. 

Let
\begin{equation}
V=\,\xi(t,x,\dot{x})\,\frac{\partial}{\partial t}+\,\eta(t,x,\dot{x})\,\frac{\partial}{\partial x}
\end{equation}
be a telescopic vector field and $\lambda$ be an arbitrary smooth function. The $\lambda$-prolongation of order $2$ of $V$, denoted by $V^{[\lambda,(2)]}$ is defined by \cite{Mur4,pucci}
\begin{equation}
V^{[\lambda,(2)]}=\,\xi(t,x,\dot{x})\,\frac{\partial}{\partial t}+\,\eta(t,x,\dot{x})\,\frac{\partial}{\partial x}+\,\eta^{[\lambda,(1)]}(t,x,\dot{x})\,\frac{\partial}{\partial \dot{x}}+\,\eta^{[\lambda,(2)]}(t,x,\dot{x})\,\frac{\partial}{\partial \ddot{x}},
\end{equation}
where
\begin{subequations}
\label{tle}
\begin{eqnarray}
\eta^{[\lambda,(1)]}(t,x,\dot{x})&=&(D+\lambda)\eta(t,x,\dot{x})-(D+\lambda)\,\xi(t,x,\dot{x})\,\dot{x},\label{tel_fir_pro}\\
\eta^{[\lambda,(2)]}(t,x,\dot{x})&=&(D+\lambda)\eta^{[\lambda,(1)]}(t,x,\dot{x})-(D+\lambda)\,\xi(t,x,\dot{x})\,\phi.\label{tel_sec_pro}
\end{eqnarray}
\end{subequations}
One may observe that without $\lambda$, the prolongation becomes usual prolongation of contact symmetries.

The invariance of the given ODE under telescopic vector field is given by \cite{Mur4}
\begin{equation}
\xi\,\frac{\partial \phi}{\partial t}+\,\eta\,\frac{\partial\phi}{\partial x}+\,\eta^{[\lambda,(1)]}\,\frac{\partial\phi}{\partial \dot{x}}-\,\eta^{[\lambda,(2)]}=\,0,\label{teles_inva}
\end{equation}
where $\eta^{[\lambda,(1)]}$ and $\eta^{[\lambda,(2)]}$ are given in Eq.(\ref{tle}).

For computational purpose we invert the expression (\ref{tel_fir_pro}) to obtain 
\begin{equation}
\lambda=\,\frac{\eta^{[\lambda,(1)]}-D[\eta]+\dot{x}\,D[\xi]}{\eta-\dot{x}\xi},
\end{equation}
so that the second prolongation can now be rewritten as
\begin{equation}
\eta^{[\lambda,(2)]}=\,D[\eta^{[\lambda,(1)]}]-\ddot{x}D[\xi]+\,\frac{\eta^{[\lambda,(1)]}-D[\eta]+\dot{x}D[\xi]}{\eta-\dot{x}\xi}(\eta^{[\lambda,(1)]}-\ddot{x}\xi).\label{second}
\end{equation}
The unknowns to be solved in (\ref{teles_inva}) now become $(\xi,\eta,\eta^{[\lambda,(1)]})$ instead of $(\xi,\eta,\lambda)$ since we have expressed $\lambda$ in terms of $(\xi,\eta,\eta^{[\lambda,(1)]})$. Substituting the given equation of motion in (\ref{teles_inva}) and solving it, we can get the components ($\xi$, $\eta$, $\eta^{[\lambda,(1)]}$) of a telescopic vector field. These telescopic vector fields are as useful as standard symmetries for what concerns symmetry reduction of an ODE \cite{gae}.

However, for a nonlinear non-polynomial oscillator (\ref{eq1}) it is very tedious to solve the determining equations arising from (\ref{teles_inva}). With the help of the procedure outlined in this paper one can obtain the telescopic vector fields for the given ODE in a simple and straightforward manner.
\section{A framework to derive adjoint-symmetries, contact symmetries and telescopic vector fields}
In this section, we present the procedure to obtain adjoint-symmetries, contact symmetries and telescopic vector fields from $\lambda$-symmetry approach. To begin, we recall the connection between $\lambda$-symmetries and integrating factors which are established through the following two equations \cite{MR09}, namely
\begin{equation}
I_x+\lambda I_{\dot x}=0,\;\;I_t+\dot x I_x+\phi I_{\dot x}=0.
\label{beq11}
\end{equation}
Differentiating both the equations with respect to $\dot x$, we obtain
\begin{eqnarray}
I_{x\dot x}+\lambda_{\dot x}I_{\dot x}+\lambda I_{\dot x \dot x}=0,~~
I_{t\dot x}+I_x+\dot x I_{x\dot x}+\phi_{\dot x}I_{\dot x}+\phi I_{\dot x \dot x}=0.
\label{beq12}
\end{eqnarray}
Defining $\mu=I_{\dot x}$ and substituting $I_x=-\lambda \mu$ in Eq.(\ref{beq12}) we find that these two expressions can be rewritten in terms of $\mu$ of the form
\begin{subequations}
\begin{eqnarray}
 \mu_x+\lambda \mu_{\dot x}+\lambda_{\dot x} \mu&=&0,\label{inte_con}\\
D[\mu]+(\phi_{\dot x}-\lambda)\mu&=&0.\label{mu_lamb}
\label{met10}
\end{eqnarray}
\end{subequations}
In the above, the function $\lambda$ is the solution of the differential equation $D[\lambda]+\lambda^2-\lambda \phi_{\dot{x}}-\phi_x=0$. Differentiating Eq.(\ref{mu_lamb}) one more time with respect to $t$ and substituting the $\lambda$-determining equation in the latter, we end up at $D^2[\mu]+D[\phi_{\dot{x}}\mu]-\phi_x \mu=0$. This expression exactly coincides with the determining equation for adjoint symmetries of the given equation \cite{Bluman}. Thus the solutions coming out from (\ref{mu_lamb}) act as adjoint-symmetries for the given equation.

Eq.(\ref{inte_con}) represents the condition for the adjoint symmetry to be an integrating factor \cite{Bluman}. We note here that the function $\mu$ comes out from Eq.(\ref{mu_lamb}) need not satisfy Eq.(\ref{inte_con}). If it satisfies the first equation for some $\lambda$ then $\mu$ becomes an integrating factor for the given ODE. If it does not satisfy (\ref{inte_con}) then it acts just as an adjoint-symmetry for the given equation. For convenience we rewrite Eq.(\ref{mu_lamb}) in the form
\begin{equation}
\lambda=\frac{D[\mu]}{\mu}+\phi_{\dot x}.
\label{beq14}
\end{equation}
From Eq.(\ref{beq14}) we can generate various symmetries. Here after we divide our analysis into two cases. In the first case we assume that the function $\lambda$ is a null function and in the second case the function $\lambda$ is a non-zero quantity. In the following, we analyze each one of the cases separately. 

\subsubsection{Case I: $\lambda=0$}
In this case, we have
\begin{equation}
 D[\mu]=-\phi_{\dot{x}}\mu.\label{s0_int}
\end{equation}
Eq.(\ref{s0_int}) exactly coincides with the determining equation for the Jacobi last multiplier for the given dynamical system \cite{nuc1}. As a consequence the solutions coming out from (\ref{s0_int}) can be considered as Jacobi last multipliers. The solutions do not satisfy (\ref{inte_con}). In other words they are not integrating factors for the given ODE. 

Once Jacobi last multiplier is known we can construct the Lagrangian for the given equation through the expression \cite{nuc2,partha}
\begin{equation}
M=\frac{\partial^2 L}{\partial \dot{x}^2}.\label{m_lag}
\end{equation}
Substituting the known expression $M=\mu$ into (\ref{m_lag}) and integrating the resultant equation two times with respect to $\dot{x}$, we obtain
\begin{equation}
 L=\int \bigg(\int M d\dot{x}\bigg)d\dot{x}+h_1(t,x)\dot{x}+h_2(t,x),\label{lagr_jlm}
\end{equation}
where $h_1$ and $h_2$ are the gauge functions of $t$ and $x$.

Interestingly once Jacobi last multiplier is obtained we can unambiguously determine the Darboux polynomials through the identity \cite{suba1}
\begin{equation}
 M=\frac{1}{F},
\end{equation}
provided the Darboux polynomial $F$ has the same cofactor $\phi_{\dot{x}}$. 
Thus, we can generate Jacobi last multipliers and Darboux polynomials from the $\lambda$-symmetry approach itself. 

\subsubsection{Case II: $\lambda \neq 0$}
Now we consider $\lambda$ is a non-zero function and is determined already for the given equation. In this case we can generate contact symmetries and telescopic vector fields for the given nonlinear ODE. More importantly we can determine the telescopic vector fields without solving the invariance condition (\ref{teles_inva}). For this purpose, we recall the expression which interrelates the telescopic vector field with $\lambda$-symmetries \cite{Mur4}, that is
\begin{eqnarray}
 \lambda=\frac{\eta^{[\lambda,(1)]}-\xi \phi}{\eta-\dot{x}\xi}.\label{lamds}
\end{eqnarray}
Here $\lambda$ is known and ($\xi,\eta,\eta^{[\lambda,(1)]}$) are the unknowns which in turn compose the telescopic vector field. Since Eq.(\ref{lamds}) has three unknowns, it is difficult to obtain the exact expressions of them just from the knowledge of $\lambda$. To overcome this obstacle we consider $\xi=0$ so that the above equation reduces to 
\begin{equation}
 \lambda=\frac{\eta^{[\lambda,(1)]}}{\eta}.\label{dumm}
\end{equation}
Recalling the relation $\lambda=-\frac{I_x}{I_{\dot{x}}}$ and substituting $\mu=I_{\dot{x}}$ into it we find $\lambda=-\frac{I_x}{\mu}$. Comparing the latter expression with (\ref{dumm}) we can rewrite Eq.(\ref{dumm}) in the form 
\begin{eqnarray}
 \lambda=\frac{\eta^{[\lambda,(1)]}}{\mu}
\end{eqnarray}
or equivalently 
\begin{equation}
 \eta^{[\lambda,(1)]}=\mu \lambda.\label{tele_zeta}
\end{equation}
Since $\mu$ and $\lambda$ are already determined the product of them now yields the third unknown $\eta^{[\lambda,(1)]}$. The other two components are $\xi=0$ and $\eta=\mu$. Thus from the knowledge of $\lambda$ we can determine the components of a telescopic vector field. In this way, from Eq.(\ref{beq11}) we can generate adjoint-symmetries, contact symmetries and the telescopic vector fields for the given second-order ODE. 
\section{Example: Mathews-Lakshmanan Oscillator}

In this section, we demonstrate the applicability of our procedure by considering Mathews-Lakshmanan oscillator Eq.(\ref{eq1}) as an example.
\subsection{Case I: Determination of Jacobi last multiplier and Darboux polynomials}
As we mentioned in Sec. II, solving Eq.(\ref{s0_int}) we can determine the Jacobi last multiplier of Eq.(\ref{eq1}). To solve Eq.(\ref{s0_int}) we assume an ansatz for $\mu$ in the form
\begin{equation}
 \mu=\frac{1}{a(t,x)+b(t,x)\dot{x}+c(t,x)\dot{x}^2},\label{ansatz}
\end{equation}
where $a(t,x),~b(t,x)$ and $c(t,x)$ are arbitrary functions of $t$ and $x$ which are to be determined. Substituting (\ref{ansatz}) into Eq.(\ref{met10}) and equating the coefficients of various powers of $\dot{x}$ to zero, we obtain a set of partial differential equations for the variables $a$, $b$ and $c$. Solving them, we obtain two particular solutions which are of the form
\begin{eqnarray}
\mu_1=\frac{1}{k  x^2+1}~~\mathrm{and}~~
\mu_2=\frac{1}{k  \dot{x}^2-\alpha^2}. \label{r1r2_ml}
\end{eqnarray}
The functions $\mu_1(=M_1)$ and $\mu_2(=M_2)$ act as Jacobi last multipliers for the nonlinear oscillator Eq.(\ref{eq1}). Substituting the Jacobi last multiplier $M_1$ in (\ref{m_lag}) and integrating it twice with respect to $\dot{x}$ and choosing the arbitrary functions which arise in this integration process as zero, we obtain the following Lagrangian for Eq.(\ref{eq1}), namely
\begin{equation}
 L_1=\frac{\dot{x}^2-\alpha^2 x^2}{2 \left(k  x^2+1\right)}.\label{lag1}
\end{equation}
The Lagrangian (\ref{lag1}) was derived by Mathews and Lakshmanan \cite{ml11,ml22} and later used by several authors \cite{sir1,recent}.

Repeating the above procedure for the second multiplier $M_2$, we find
\begin{eqnarray}
 L_2=\frac{\alpha \log \left(\frac{k  x^2+1}{\alpha^2-k  \dot{x}^2}\right)-2 \sqrt{k } \dot{x} \tanh ^{-1}\left(\frac{\sqrt{k } \dot{x}}{\alpha}\right)}{2 k  \alpha}.
\end{eqnarray}
where we have chosen the gauge term as
\begin{eqnarray}
 h_2=\frac{\log(1+k x^2)}{2k}.
\end{eqnarray}

The Darboux polynomials can be unambiguously derived from the Jacobi last multipliers through the identity $F=\frac{1}{M}$. The respective Darboux polynomials are 
\begin{eqnarray}
F_1=k  x^2+1~~\mathrm{and}~~F_2=k  \dot{x}^2-\alpha^2.
\end{eqnarray}
Both of them share the same cofactor, namely
\begin{equation}
 g_1=g_2=\frac{2 k  \dot{x} x}{k  x^2+1}.
\end{equation}
The ratio of two Darboux polynomials which share the same cofactor provide the first integral \cite{suba1}. From $F_1$ and $F_2$, we find
\begin{equation}
 I=\frac{F_2}{F_1}=\frac{k\dot{x}^2-\alpha^2}{1+k x^2}.\label{first_inte}
\end{equation}
The integral can be interpreted as the Hamiltonian for the nonlinear oscillator Eq.(\ref{eq1}).
\subsection{Case II: Determination of other symmetries} 
Now we assume $\lambda \neq 0$ in Eq.(\ref{mu_lamb}) and solve the underlying equation (\ref{mu_lamb}) and obtain adjoint-symmetries, contact symmetries and telescopic vector fields of Eq.(\ref{eq1}). 
\subsubsection{Adjoint-symmetries}
To solve Eq.(\ref{mu_lamb}) we should know the $\lambda$-function. As our aim is to derive adjoint-symmetries, contact symmetries and telescopic vector fields from $\lambda$-symmetries and not to determine the later, we consider the same $\lambda$-functions which are reported by Bhuvaneshwari et.al. \cite{bhu_ml}, that is
\begin{subequations}
\label{lam}
 \begin{eqnarray}
 \lambda_1&=&-\frac{\left(k  \dot{x}^2-\alpha^2\right) \left(k  t \alpha^2 x^3+k  t x \dot{x}^2+k  x^2 \dot{x}+\dot{x}\right)}{\left(k  x^2+1\right) \left(\alpha^2 x-k  \left(t \alpha^2 x^2 \dot{x}+t \dot{x}^3-\alpha^2 x^3\right)\right)},\label{lam1}\\
 \lambda_2&=&\frac{x}{\dot x}\bigg(\frac{k{\dot x}^2-\alpha^2}{1+kx^2}\bigg).
\label{lam2}
\end{eqnarray}
\end{subequations}
The $\lambda$-symmetry vector field is given by $v=\frac{\partial}{\partial x}$. 

We substitute these two $\lambda$-functions separately into (\ref{mu_lamb}) and solve the resultant equations to obtain the function $\mu$. To solve the $\mu$- determining equation (\ref{mu_lamb}) we assume an ansatz for $\mu$ in the form
\begin{equation}
\mu=\frac{a_1(t,x)+a_2(t,x)\dot{x}+a_3(t,x)\dot{x}^2+a_4(t,x)\dot{x}^3}{b_1(t,x)+b_2(t,x)\dot{x}+b_3(t,x)\dot{x}^2},
\end{equation}
where $a_i's,~~i=1,2,3,4$ and $b_j's,~j=1,2,3$, are arbitrary functions of $t$ and $x$. Substituting this ansatz into (\ref{mu_lamb}) and equating the coefficients of $\dot{x}$ and solving the resultant equations we obtain  two particular solutions which are of the form
\begin{subequations}
\label{inte_r}
\begin{eqnarray}
  \mu_1&=&\frac{\alpha^2 x \left(-k  t x \dot{x}+k  x^2+1\right)-k  t \dot{x}^3}{\left(k  x^2+1\right) \left(\alpha^2 x^2+\dot{x}^2\right)},\\
  \mu_2&=&-\frac{2 k  \dot{x}}{k  x^2+1}\label{r1}.
\end{eqnarray}
\end{subequations}
Among these two, $\mu_2$ acts as an integrating factor and adjoint-symmetry for Eq.(\ref{eq1}) since it satisfies both the expressions (\ref{inte_con}) and (\ref{mu_lamb}). On the other hand $\mu_1$ acts only as an adjoint-symmetry for (\ref{eq1}) since it does not satisfy Eq.(\ref{inte_con}). 
\subsubsection{Contact symmetries}
Now we determine contact symmetries admitted by Eq.(\ref{eq1}). This can be done by exploiting the relation $\lambda=\frac{D[Q]}{Q}$, where $Q$ is the characteristics and is given by $Q=\eta-\dot{x}\xi$. Here we assume that the functions $\eta$ and $\xi$ are functions of $t$, $x$ and $\dot{x}$. Substituting the known function $\lambda$ into this expression and assuming a suitable ansatz for $Q$ and substituting this ansatz on the right hand side and solving the resultant equations we can obtain the exact form of $Q$ from which we can identify point/contact symmetries. 

To determine the characteristics $Q$ we consider the following ansatz for $Q$:
\begin{equation}
 Q=\frac{a(t,x)+b(t,x)\dot{x}+c(t,x)\dot{x}^2+d(t,x)\dot{x}^3}{e(t,x)},
\end{equation}
where $a,b,c,d$ and $e$ are arbitrary functions which are all to be determined. We start our analysis with $\lambda_1$. We substitute the above form of $Q$ and its total derivative and the function $\lambda_1$ (see Eq.(\ref{lam1})) in the relation $\lambda=\frac{D[Q]}{Q}$ and implement the above procedure. After a very lengthy calculation we find a particular solution which is of the form
\begin{eqnarray}
 Q_1=\frac{\alpha^2 x \left(-k  t x \dot{x}+k  x^2+1\right)-k  t \dot{x}^3}{k  x^2+1}.
\end{eqnarray}
From the definition $Q=\eta-\dot{x}\xi$, we identify the functions $\xi$ and $\eta$ are to be
\begin{eqnarray}
 \xi=\frac{-k  t \dot{x}^2}{k  x^2+1} ~~\mathrm{and}~~\eta=\frac{\alpha^2 x \left(-k  t x \dot{x}+k  x^2+1\right)}{k  x^2+1}.\label{cont_ex}
\end{eqnarray}
The functions $\xi$ and $\eta$ confirm that the underlying symmetry is a contact one. The associated vector field is given by
\begin{eqnarray}
 \Omega_1=\frac{-k  t \dot{x}^2}{k  x^2+1}\frac{\partial}{\partial t}+\frac{\alpha^2 x \left(-k  t x \dot{x}+k  x^2+1\right)}{k  x^2+1}\frac{\partial}{\partial x}.
\end{eqnarray}
One can check that the functions $\xi$ and $\eta$ given in (\ref{cont_ex}) satisfy the contact symmetry invariance condition $\xi \frac{\partial \phi}{\partial t}+\eta \frac{\partial \phi}{\partial x}+\eta^{(1)}\frac{\partial \phi}{\partial \dot{x}}-\eta^{(2)}=0$, where $\eta^{(1)}=D[\eta]-\dot{x}D[\xi]$ and $\eta^{(2)}=D[\eta^{(1)}]-\phi D[\xi]$. 

Repeating the procedure for the function $\lambda_2$, given in (\ref{lam2}), we find $
 Q=-\dot{x}$. Since the above expression is linear in $\dot{x}$, the functions $\xi$ and $\eta$ should be of point-type, that is $\xi=1$ and $\eta=0$ which is nothing but the time translational symmetry. Since the integrals and the general solution of (\ref{eq1}) are already known we do not derive them from the contact symmetry given above.


\subsubsection{Telescopic vector fields}
Substituting the known $\lambda$-functions (\ref{lam}) and the adjoint-symmetries $\mu$ (\ref{inte_r}) separately into Eq.(\ref{tele_zeta}) we can obtain the components of the telescopic vector field as
\begin{subequations}
\begin{eqnarray}
&&(i)~~\xi_1=0,~~\eta_1=\frac{\alpha^2 x (k  x (x-t \dot{x})+1)-k  t \dot{x}^3}{\left(k x^2+1\right) \left(\alpha^2 x^2+\dot{x}^2\right)},\nonumber \\
&&  \eta_1^{[\lambda,(1)]}=\frac{\left(\alpha^2-k  \dot{x}^2\right) \left(k  t \alpha^2 x^3+k  x \dot{x} (t \dot{x}+x)+\dot{x}\right)}{\left(k  x^2+1\right)^2 \left(\alpha^2 x^2+\dot{x}^2\right)},\\
&&(ii)~~  \xi_2=0,~~\eta_2=-\frac{2 k  \dot{x}}{k  x^2+1},~~
  \eta_2^{[\lambda,(1)]}=-\frac{2 k  x \left(k  \dot{x}^2-\alpha^2\right)}{\left(k  x^2+1\right)^2}.
\end{eqnarray}
\end{subequations}
The associated telescopic vector fields are given by
\begin{subequations}
\begin{eqnarray}
 V_1&=&\frac{\alpha^2 x (k  x (x-t \dot{x})+1)-k  t \dot{x}^3}{\left(k x^2+1\right) \left(\alpha^2 x^2+\dot{x}^2\right)}\frac{\partial}{\partial x}+\frac{\left(\alpha^2-k  \dot{x}^2\right) \left(k  t \alpha^2 x^3+k  x \dot{x} (t \dot{x}+x)+\dot{x}\right)}{\left(k  x^2+1\right)^2 \left(\alpha^2 x^2+\dot{x}^2\right)}\frac{\partial}{\partial \dot{x}}\nonumber\\
&&+\frac{\left(\alpha^2-k  \dot{x}^2\right) \left(\alpha^2 x \left(k  t x \dot{x}+k  x^2-1\right)+k  \dot{x}^2 (t \dot{x}+2 x)\right)}{\left(k x^2+1\right)^2 \left(\alpha^2 x^2+\dot{x}^2\right)}\frac{\partial}{\partial \ddot{x}},\\
 V_2&=&-\frac{2 k  \dot{x}}{k  x^2+1}\frac{\partial}{\partial x}-\frac{2 k  x \left(k  \dot{x}^2-\alpha^2\right)}{\left(k  x^2+1\right)^2}\frac{\partial}{\partial \dot{x}}-\frac{2 k  \dot{x} \left(k  \dot{x}^2-\alpha^2\right)}{\left(k  x^2+1\right)^2}\frac{\partial}{\partial \ddot{x}}.
\end{eqnarray}
\end{subequations}

One can check that the above telescopic vector fields satisfy the invariance condition (\ref{teles_inva}). In vector fields $V_1$ and $V_2$ the expressions for the second prolongation  $\eta^{[\lambda,(2)]}$ are found through Eq.(\ref{second}). 

\section{Conclusion}
In this paper, we have proposed a method of deriving adjoint-symmetries, contact symmetries, telescopic vector fields, Jacobi last multiplier and Darboux polynomial for a given second-order nonlinear ODE. We have shown that all these quantities can be obtained from the $\lambda$-symmetry approach itself. The determining equations for the telescopic vector fields are in general tedious to solve. Under this circumstance the procedure proposed here may be adopted to explore the telescopic vector fields for the given nonlinear dynamical system. Eventhough the applicability of this procedure has been demonstrated by considering the nonlinear non-polynomial oscillator (\ref{eq1}), the underlying methodology can be extended to a class of second-order nonlinear oscillator equations. We have also verified the above said procedure with another example which was discussed by Ballesteros et.al.\cite{ball}. For this equation we have derived adjoint-symmetries, contact symmetries, telescopic vector fields, Darboux polynomials and Jacobi last multipliers. Since the procedure is exactly the same we do not present the results here. 
\section*{Acknowledgments}
RMS acknowledges the University Grants Commission (UGC-RFSMS), Government of India, for providing a Research Fellowship. The work of MS forms part of a research project sponsored by Department of Science and Technology, Government of India.

\end{document}